# Determining Graphene Adhesion via Substrate-regulated Morphology of Graphene

Zhao Zhang[a], Teng Li[a,b,*]

[a]Department of Mechanical Engineering, University of Maryland, College Park, MD 20742

[b]Maryland NanoCenter, University of Maryland, College Park, MD 20742

**Abstract**

Understanding the adhesion between graphene and other materials is crucial for achieving more reliable graphene-based applications in electronic devices and nanocomposites. The ultra-thin profile of graphene, however, poses significant challenge to direct measurement of its adhesion property using conventional approaches. We show that there is a strong correlation between the morphology of graphene on a compliant substrate with patterned surface and the graphene-substrate adhesion. We establish an analytic model to quantitatively determine such a strong correlation. Results show that, depending on the graphene-substrate adhesion, number of graphene layers and substrate stiffness, graphene exhibits two distinct types of morphology: I) graphene remains bonded to the substrate and corrugates to an amplitude up to that of the substrate surface patterns; II) graphene debonds from the substrate and remains flat on top of the substrate surface patterns. The sharp transition between these two types of graphene morphology occurs at a critical adhesion between the graphene and the compliant substrate material. These results potentially open up a feasible pathway to measuring the adhesion property of graphene.

---

[*] Author to whom correspondence should be addressed. Electronic mail: LiT@umd.edu (TL)

## I. INTRODUCTION

The exceptional electronic and mechanical properties of graphene[1,2] have inspired tantalizing potential applications, such as transparent flexible displays,[3] biochemical sensing arrays[4] and graphene-reinforced nanocomposites.[5] Since the electronic properties of graphene are closely tied to its morphology, controlling graphene morphology over large areas becomes essential in enabling future graphene-based devices.[6,7] Moreover, in order to achieve more reliable graphene-based electronics and graphene-reinforced nanocomposites, it is crucial to understand the adhesion between graphene and other materials (e.g., a flexible substrate or a polymer matrix). However, direct measurement of the adhesion property of ultra-thin graphene is rather challenging,[8] as the traditional metrology of adhesion at macroscopic scales becomes unsuitable in dealing with samples of extremely small dimension. For example, the commonly used peeling tests, wedge tests, or double-cantilever beam methods require precise determination of the interfacial cracking, which becomes extremely challenging in manipulating ultra-thin films such as graphene. In this paper, we present an analytic model to explicitly determine the morphology of mono-layer and few-layer graphene regulated by an underlying compliant elastic substrate with patterned surface, and show that such regulated morphology of graphene is strongly dependent on the graphene-substrate adhesion. In particular, the sharp transition between two distinct types of graphene morphology on the patterned surface of the compliant substrate can be used to quantitatively determine the graphene-substrate adhesion. Results from this study, on the one hand offer quantitative guideline for controlling graphene morphology on compliant substrates, and on the other hand open up a feasible pathway to characterizing the adhesion properties between graphene and various elastic materials.

Recent experiments show that the morphology of substrate-supported graphene is largely regulated by the substrate surface, distinct from the random corrugations of freestanding graphene.[9-13] For example, mono-layer and few-layer graphene are shown to partially follow the surface morphology of various substrates (e.g., GaAs, InGaAs and $SiO_2$).[11-14] These observations have motivated analytic and computational models to quantitatively determine the regulated graphene morphology.[15-20] However, the existing analytic and computational models are mainly based on two assumptions. First, the substrate is assumed to be rigid and thus does not deform when interacting with the graphene. Second, only mono-layer graphene is considered. Results from these existing models shed important light on the substrate-regulated graphene morphology; however, the two assumptions limit the general applicability of these models. In reality, it is much easier to fabricate few-layer graphene than mono-layer graphene, and thus few-layer graphene is more commonly used in applications such as graphene-reinforced nanocomposites. Transfer printing technique also allows for transferring graphene from a mother wafer onto a wide range of substrate materials, such as polymers and elastomers.[21,22] The morphology of few-layer graphene regulated by a compliant substrate depicts rather rich characteristics that cannot be readily captured by the existing models. For example, recent experiments show that an 8-layer graphene on a compliant polydimethylsiloxane (PDMS) substrate can closely conform to the sinusoidal surface grooves of the PDMS. By contrast, a 13-layer graphene remains nearly flat on the grooved substrate surface.[22] Although a theoretical model is offered to calculate the deformation in the graphene-substrate system, the bending rigidity of few-layer graphene is over-estimated and the tension in graphene is neglected.[22] To overcome the limitations of existing models, we establish a generalized analytic model to explicitly determine the morphology of *mono-layer* and *few-layer* graphene regulated by the patterned surface of a *compliant* elastic substrate.

## II. ANALYTIC MODEL

Figure 1 illustrates the typical transfer printing process of graphene from a mother wafer to a compliant substrate with patterned surface and possible resulting structures. An *n*-layer graphene ($n \geq 1$) fabricated on a stiff and smooth mother wafer (e.g., silica for mechanically exfoliated graphene or copper for chemically grown graphene) is brought in contact with a compliant substrate (e.g., polymer or elastomer) with patterned surface, pressure is then applied to guarantee the full contact between the graphene and the substrate (e.g., the patterned surface is flattened under pressure). Upon release of the pressure, the mother wafer is lifted from the compliant substrate. If the graphene adheres more strongly to the compliant substrate than to the mother wafer, the graphene is left on the compliant substrate.

The resulting morphology of the graphene on the patterned surface of the compliant substrate is dictated by the competition between the graphene-substrate adhesion energy and the strain energy in the graphene-substrate laminate. The regulated graphene morphology can be categorized into two types:

**Type I**: If the graphene-substrate adhesion energy (denoted as $\Gamma_{gs}$) is strong, the graphene remains bonded to the compliant substrate at the price of increased strain energy due to the corrugation of the graphene (denoted as $E_g$) and the distortion of the substrate near the portion underneath the graphene (denoted as $E_s$) (e.g., Figs. 1b-d). The amplitude of the graphene corrugation can be determined through minimizing the total free energy (i.e., $E_g + E_s - \Gamma_{gs}$), as illustrated in Fig. 1e;

**Type II**: If the graphene-substrate adhesion energy is weak and cannot balance the aforementioned strain energy of the graphene-substrate laminate (i.e., $E_g+E_s > \Gamma_{gs}$, Fig. 1g), the graphene-substrate interface debonds. As a result, the graphene remains nearly flat on the top of the patterned substrate surface while the substrate surface recovers to its original pattern (Fig. 1f). In such a case, the strain energy of the graphene-substrate laminate is negligible.

After the transfer printing process, the strain energy of the Type I corrugated graphene consists of the contributions from bending and stretching of the graphene. While the bending energy of the graphene is determined by its out-of-the-plane deflection, the membrane energy of the graphene depends on both its in-plane displacement and out-of-plane deflection. In reality, relative sliding between the graphene and the underlying substrate may occur during transfer printing, which can mitigate the in-plane stretching of the graphene. Such relative sliding depends on the graphene-substrate friction and detailed transfer printing conditions, which is often difficult to quantify. To overcome such an uncertainty, here we consider the following two limiting cases.

In one limiting case, we assume there is no relative sliding, i.e., the graphene deforms from a flat profile to a sinusoidal wavy profile by purely deflecting out of the plane while the in-plane displacement of the graphene is zero. Assuming the profiles of the patterned substrate surface and the Type I corrugated graphene morphology in *x-y* plane to be $w_s(x,y)$ and $w_g(x,y)$, respectively, the bending energy of the graphene is given by

$$E_g^b = \iint \frac{D}{2}\left[\left(\frac{\partial^2 w_g}{\partial^2 x}+\frac{\partial^2 w_g}{\partial^2 y}\right)^2 + 2(1-\nu)\left(\left(\frac{\partial^2 w_g}{\partial x \partial y}\right)^2 - \frac{\partial^2 w_g}{\partial^2 x}\frac{\partial^2 w_g}{\partial^2 y}\right)\right] dx dy \quad (1)$$

where *D* is the bending rigidity and $\nu$ the Poisson's ratio of the graphene, respectively. The in-plane strain of the graphene in such a limiting case results from the out-of-plane deflection and the strain components are given by

$$\varepsilon_{xx} = \frac{1}{2}\left(\frac{\partial w_g}{\partial x}\right)^2, \varepsilon_{yy} = \frac{1}{2}\left(\frac{\partial w_g}{\partial y}\right)^2 \text{ and } \varepsilon_{xy} = \frac{1}{2}\frac{\partial w_g}{\partial x}\frac{\partial w_g}{\partial y}. \tag{2}$$

Therefore the membrane energy of the graphene is given by

$$E_g^m = \iint \frac{C}{2}\left[(\varepsilon_{xx} + \varepsilon_{yy})^2 + 2(1-\nu)(\varepsilon_{xy}^2 - \varepsilon_{xx}\varepsilon_{yy})\right]dxdy \tag{3}$$

where $C$ is the in-plane elastic modulus of the graphene. The total strain energy of the graphene is thus given by

$$E_g = E_g^b + E_g^m \tag{4}$$

In another limiting case, the graphene is allowed to slide freely on the substrate surface so that the stretching in the graphene can be fully relaxed. In other words, the strain energy of the graphene results solely from its bending, that is,

$$E_g = E_g^b \tag{5}$$

Assuming the distortion deformation of the compliant substrate underneath the graphene is elastic, the resulting strain energy of the substrate is equivalent to the work done by the graphene-substrate interfacial traction $p(x,y)$ over the distortion displacement of the substrate surface $(w_s - w_g)$. That is,

$$E_s = \iint \frac{1}{2}p(w_s - w_g)dxdy \tag{6}$$

The total strain energy, $E_g + E_s$, obtained from the above formulation can then be compared with the graphene-substrate adhesion energy, $\Gamma_{gs}$, to determine the resulting graphene morphology. For example, for a given substrate material and its surface profile, $E_g + E_s$ is computed over a certain range of graphene corrugation amplitude (e.g., from zero to the amplitude of the substrate surface pattern). If $(E_g + E_s)_{min} < \Gamma_{gs}$, the graphene remains bonded with the substrate and corrugate with an amplitude corresponding to $(E_g + E_s)_{min}$ (i.e., Type I, Fig. 1e). If $(E_g + E_s)_{min} >$ $\Gamma_{gs}$, the graphene debonds from the substrate and remains flat on the substrate grooves (i.e., Type II, Fig. 1g).

## III. RESULTS AND DISCUSSION

We next apply the above analytic model formulation to determine the morphology of *n*-layer graphene regulated by a compliant substrate patterned with sinusoidal surface grooves. The profiles of the surface grooves and the resulting Type I graphene morphology are described by $w_s = A_s\cos(2\pi x/\lambda)$ and $w_g = A_g\cos(2\pi x/\lambda)$, respectively, where $A_s$ and $A_g$ are the amplitude of the grooves and the graphene corrugation, respectively, and $\lambda$ the wavelength. Substituting $w_g$ into Eqs. (1)-(3) gives that, for the limiting case of no relative sliding of the graphene, the average strain energy of the graphene over one groove period is

$$E_g = \frac{1}{\lambda}\int_0^\lambda \left[\frac{D}{2}\left(\frac{\partial^2 w_g}{\partial^2 x}\right)^2 + \frac{C}{8}\left(\frac{\partial w_g}{\partial x}\right)^4\right]dx = \frac{4\pi^4 D A_g^2}{\lambda^4} + \frac{3\pi^4 C A_g^4}{4\lambda^4}, \quad (7)$$

where the first term on the right side denotes the contribution from the bending energy and the second the contribution from the membrane energy; and thus for the limiting case of graphene freely sliding on the substrate surface, the average strain energy of the graphene over one groove period is

$$E_g = \frac{4\pi^4 D A_g^2}{\lambda^4}. \quad (8)$$

The distortion deformation of the elastic substrate surface underneath the graphene $(A_s - A_g)\cos(kx)$ results in a graphene-substrate interfacial traction $p = (\pi \tilde{Y}_s/\lambda)(A_s - A_g)\cos(kx)$,[23-25] where $\tilde{Y}_s$ is the plane strain Young's Modulus of the substrate material. Thus, the average strain energy of the substrate over one groove period is given by

$$E_s = \frac{\pi \tilde{Y}_s}{4\lambda}(A_s - A_g)^2. \quad (9)$$

To benchmark the above formulation, we take $\Gamma_{gs} = 0.1\ J/m^2$ (representative of graphene-polymer adhesion), $\lambda = 1.5\ \mu m$ and $A_s = 100\ nm$ (comparable to recent experiments[22]). For an $n$-layer graphene, its bending rigidity $D$ is taken to be $(3.8n^3 - 3.6n^2) \times 10^{-18}\ Nm$,[26] and its in-plane elastic modulus $C = 340n\ N/m$.[2] Figure 2 plots the normalized graphene amplitude $A_g/A_s$ as a function of the substrate plane strain Young's modulus $\tilde{Y}_s$ for various numbers of graphene layers $n = 1$, 10 and 35, respectively. For the limiting case of no graphene sliding on the substrate (Fig. 2a), if the substrate is very compliant (e.g., $\tilde{Y}_s \leq 1 MPa$), the graphene remains bonded to the substrate and assumes a rather flat morphology (e.g., Type I, $A_g/A_s \ll 1$). In other words, the substrate surface grooves underneath the graphene is nearly flattened. As the substrate becomes stiffer, the graphene becomes more corrugated (increasing $A_g$). For a given substrate stiffness, the thicker the graphene layers (higher bending rigidity), the less the graphene is corrugated. For a given $n$, however, there exists a critical substrate stiffness, higher than which the graphene debonds from the substrate and remains flat on the substrate surface grooves (e.g., Type II, $A_g/A_s = 0$). The transition from Type I to Type II graphene morphology is sharp. The critical substrate stiffness decreases as $n$ increases. For the limiting case of the graphene freely sliding on the substrate (Fig. 2b), a monolayer graphene completely conforms to the surface grooves of a substrate of any stiffness ($A_g/A_s = 1$), a few-layer graphene (e.g., $n = 10$) corrugates slightly on a rather compliant substrate but conforms closely to the surface of a sufficiently stiff substrate. However, the morphology of a thick graphene layer can sharply switch between Type I and Type II at a critical substrate stiffness (e.g., at $\tilde{Y}_s \approx 100 MPa$ for $n = 35$).

Figure 3 further plots $A_g/A_s$ as a function of $n$ for $\tilde{Y}_s = 1 MPa, 10\ MPa$ and $1 GPa$, respectively. For the limiting case of no graphene sliding on the substrate (Fig. 3a), if the substrate is compliant (e.g., $\tilde{Y}_s = 1MPa$ or $10MPa$), the graphene remains bonded to the substrate and assumes a slightly corrugated morphology (e.g., Type I). For a given $\tilde{Y}_s$, $A_g$ decreases as $n$ increases. On a sufficiently stiff substrate (e.g., $\tilde{Y}_s = 1GPa$), graphene with any number of layers debonds from the substrate and remains flat (Type II). For the limiting case of the graphene freely sliding on the substrate (Fig. 3b), if the substrate is compliant (e.g., $\tilde{Y}_s = 1MPa$ or $10MPa$), graphene remains bonded to the substrate and $A_g$ decreases gradually as $n$ increases (Type I). If the substrate is sufficiently stiff (e.g., $\tilde{Y}_s = 1GPa$), a thinner graphene ($n \leq 32$) remains bonded and fully conformed to the substrate (Type I) while a thicker graphene ($n \geq 33$) debonds from the substrate and remains flat (Type II). Such a sharp transition in graphene morphology is similar to the snap-through instability of graphene morphology on a corrugated substrate predicted by models and observed in experiments.[16,17,22]

The sharp transition between Type I and Type II graphene morphologies shed light on characterizing the graphene-substrate adhesive properties. As an illustration, Fig. 4 maps the minimum strain energy of the graphene-substrate laminate, $(E_g + E_s)_{min}$, in the space of $\tilde{Y}_s$ and $n$, for the limiting cases of no graphene sliding (Fig. 4a) and graphene sliding freely (Fig. 4b) on the substrate surface. If, for a given $\tilde{Y}_s$, a critical number of graphene layers $n_{cr}$ can be determined from experiments at which the graphene morphology switches between Type I and Type II, the energy levels corresponding to $(n_{cr}, \tilde{Y}_s)$ in Figs. 4a and 4b define the upper and lower bounds of the graphene-substrate adhesion energy $\Gamma_{gs}$. For example, taking $n_{cr} = 13$, $\tilde{Y}_s = 1.6MPa$, Fig. 4 gives $\Gamma_{gs}$ ranging from 3.5 $mJ/m^2$ to 7.4 $mJ/m^2$, which agrees reasonably well with the experimental results on graphene-PDMS adhesion[22] (~7.1 $mJ/m^2$). On the other hand, for a given $\Gamma_{gs}$, the corresponding contour line in the energy map defines a boundary below which the graphene assumes Type I morphology and above which it assumes Type II morphology. For example, the solid contour line of $\Gamma_{gs} = 0.1\ J/m^2$ in Fig. 4a intersects the dotted lines of $n = 1$, 10, and 35, defining a critical substrate stiffness for each $n$ that corresponds to the sharp transition between Type I and Type II morphology revealed in Fig. 2a. Similarly, the solid contour line in Fig. 4b intersects the dotted line of $n = 35$ and that of $\tilde{Y}_s = 1 GPa$, defining a critical substrate stiffness and a critical number of graphene layer that correspond to the sharp morphologic transition revealed in Figs. 2b and 3b, respectively. While Fig. 4 is specifically applicable to the case of few-layer graphene on a compliant substrate with sinusoidal surface grooves, similar energy maps for few-layer graphene morphology regulated by other patterned substrate surfaces can be readily obtained following the formulation of the generalized analytic model delineated in Section II.

The present model assumes no inter-layer sliding between graphene layers during the transfer printing process. In reality, when the curvature of substrate surface is large and the graphene-substrate adhesion energy is strong, inter-layer shearing between different graphene layers due to conforming to the substrate surface may be severe enough to cause inter-layer sliding. As a result, the strain energy of the few-layer graphene is partially relaxed. In this sense, the present model overestimates the adhesion energy between the graphene and the substrate if inter-layer sliding occurs. Besides inter-layer sliding, the separation between graphene layers may also occur. When the top layers of graphene separate from the bottom ones, the strain energy in the graphene-substrate structure is partially released, serving as the driving force for inter-layer separation. The critical value of such a driving force to initiate graphene inter-layer separation is estimated to be 0.29 J/m², the carbon-carbon inter-layer adhesion energy.[27] For the limiting case of no sliding between the graphene and the substrate (i.e., Fig. 4a), for a given substrate stiffness $\tilde{Y}_s$, $(E_g+E_s)_{min}$ increases monotonically with the number of graphene layers, $n$. Therefore, there exists a critical number of graphene layers $n_{cr}$ for a given $\tilde{Y}_s$, at which the corresponding $(E_g+E_s)_{min}$ overweighs that for the case of $n = 1$ by 0.29 J/m$^2$. In other words, if the few-layer graphene is too thick (e.g., $n > n_{cr}$), inter-layer separation may occur. The dashed line in Fig. 4a plots the corresponding $n_{cr}$ for various $\tilde{Y}_s$, which estimates a boundary in the space of $\tilde{Y}_s$ and $n$, below which inter-layer separation in few-layer graphene does not occur and thus the present model is valid. For the limiting case that graphene can slide freely on the substrate surface (i.e., Fig. 4b), the total strain energy density in the graphene-substrate structure is less than 0.29 J/m$^2$. In other words, there is no sufficient driving force to initiate the inter-layer separation. Therefore, the present model is valid within the full space of $\tilde{Y}_s$ and $n$ used in Fig. 4b.

## IV. CONCLUSIONS

In summary, we show that strong correlation exists between the adhesion property of graphene and its morphology regulated by the patterned surface of a compliant substrate. We delineate an analytic model to quantitatively determine the regulated morphology of the graphene. Two distinct types of graphene morphology emerge from the results: Type I) graphene remains bonded to the substrate and corrugates to an amplitude up to that of the substrate surface patterns; Type II) graphene debonds from the substrate and remains flat on top of the substrate surface patterns. The sharp transition between these two types of graphene morphology can potentially open up a feasible pathway to characterizing the adhesion between graphene and various elastic materials, a property that is rather challenging to measure directly. We therefore call for further experiments to explore such an approach.

**ACKNOWLEDGEMENT**

This research is supported by National Science Foundation (grants #1069076 and #1129826) and a UMD GRB Summer Research Award. Z.Z. also thanks the support of A. J. Clark Fellowship, UMD Clark School Future Faculty Program, and Ellen D. Williams Distinguished Fellowship.

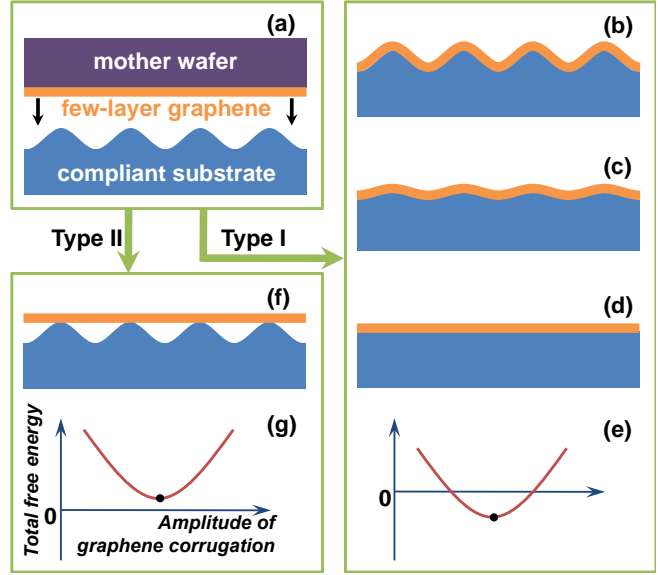

FIG 1. (Color online) (a) Schematic of the transfer printing of a few-layer graphene from a flat and stiff mother wafer onto a compliant substrate with sinusoidal surface grooves. The resulting graphene morphology can be categorized into two types. Type I: If the graphene-substrate adhesion is sufficiently strong, graphene remains bonded to the compliant substrate and corrugates to an amplitude up to that of the substrate surface grooves (b-d). The graphene amplitude depends on the substrate stiffness and the number of graphene layers. Type II: If the graphene-substrate adhesion is weak, graphene debonds from the substrate and remains flat on top of the substrate surface grooves (f). (e) and (g) schematically plot the total free energy, $E_g + E_s - \Gamma_{gs}$, as a function of the amplitude of graphene corrugation in Type I and Type II, respectively.

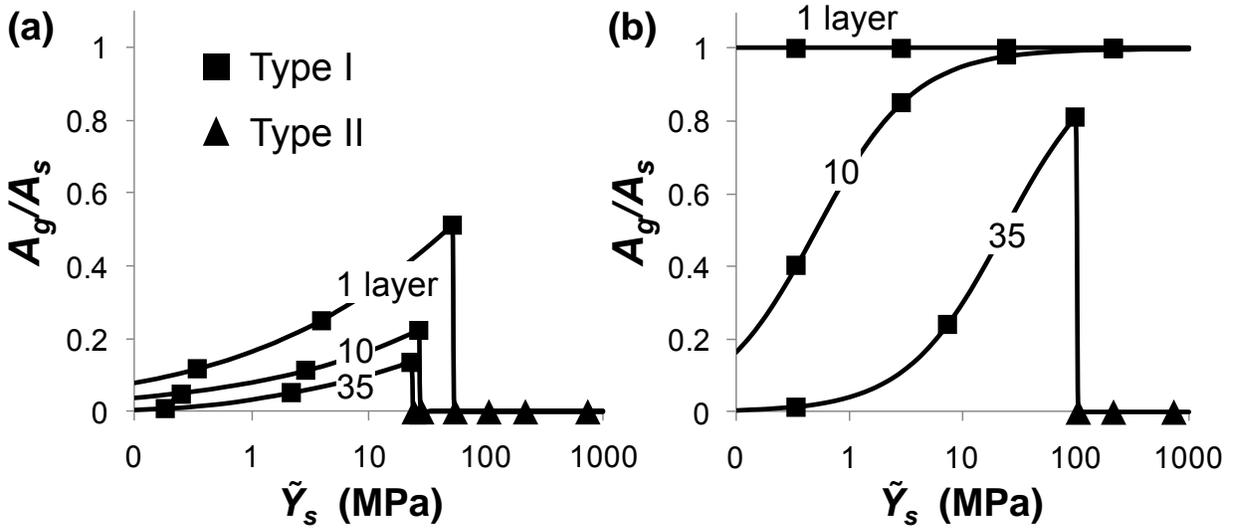

FIG. 2. The normalized graphene amplitude $A_g/A_s$ as a function of $\tilde{Y}_s$ for $n =1$, 10 and 35, respectively, for (a) the limiting case of no graphene sliding on the substrate and (b) the limiting case of graphene freely sliding on the substrate. Note the sharp transition between Type I (square marks) and Type II (triangle marks) graphene morphology at certain combinations of $\tilde{Y}_s$ and $n$.

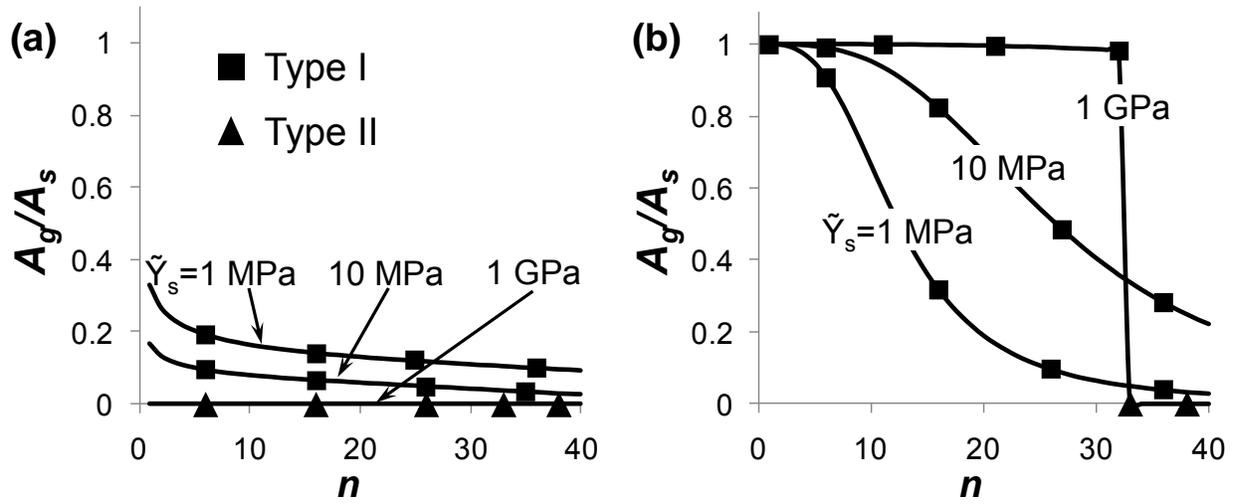

FIG. 3. The normalized graphene amplitude $A_g/A_s$ as a function of $n$ for $\tilde{Y}_s = 1$ MPa, 10 MPa and 1 GPa, respectively, for (a) the limiting case of no graphene sliding on the substrate and (b) the limiting case of graphene freely sliding on the substrate. Note the sharp transition between Type I (square marks) and Type II (triangle marks) graphene morphology at certain combinations of $\tilde{Y}_s$ and $n$.

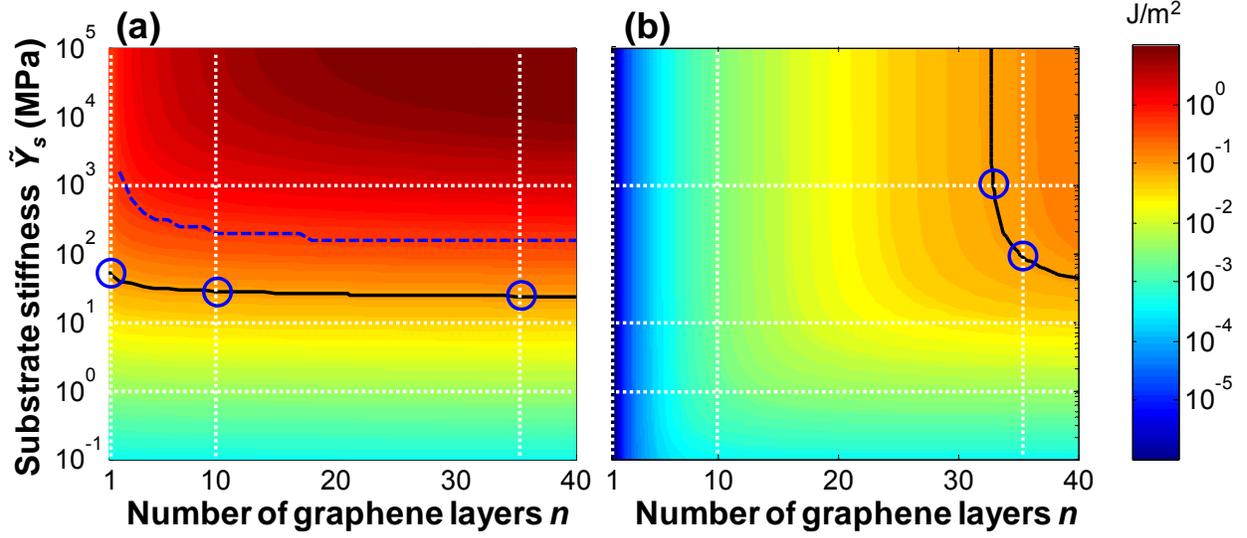

FIG. 4. (Color online) The map of $(E_g+E_s)_{min}$ in the space of $\tilde{Y}_s$ and $n$ for the limiting cases of (a) no graphene sliding and (b) graphene freely sliding on the substrate, respectively. The energy levels in (a) and (b) corresponding to a combination of $\tilde{Y}_s$ and $n$ at which graphene morphology switches between Type I and Type II define the upper and lower estimates of the graphene-substrate adhesion energy $\Gamma_{gs}$. The solid contour line denotes $\Gamma_{gs} = 0.1\,J/m^2$, which defines a boundary below which the graphene assumes Type I morphology and above which it assumes Type II morphology. The three vertical dotted lines in (a) and those in (b) correspond to the cases in Figs. 2a and b, respectively. The three horizontal dotted lines in (a) and those in (b) correspond to the cases in Figs. 3a and b, respectively. The intersections of the dotted lines and the solid lines (circles) indicate the sharp transitions between Type I and Type II in Figs. 2 and 3. For a given $\tilde{Y}_s$, the dashed line in Fig. 4a defines a critical number of graphene layers, larger than which the corresponding $(E_g+E_s)_{min}$ is 0.29 J/m² greater than that for the case of $n = 1$. Inter-layer separation of the graphene may occur in the region above this dashed line.

17